# Ethics of Technology needs more Political Philosophy

Johannes Himmelreich, Maxwell School of Citizenship and Public Affairs, Syracuse University

As a driver, have you ever asked yourself whether to make left turns? Unprotected left turns, that is, left turns with oncoming traffic, are among the most difficult and dangerous driving maneuvers. Although the risk of each individual left turn is negligible, if you are designing the behavior of a large fleet of self-driving cars, small individual risks add up to a significant number of expected injuries in the aggregate. Whether a fleet of cars should make left turns is a question that any developer of self-driving cars and any designer of mapping and routing applications faces today.

A more general issue is at stake here: the decision of whether to make left turns involves a trade-off between safety and mobility (the time it takes to get to a destination). You gain safety at the expense of mobility by driving around the block and thereby avoiding left turns. But you gain mobility at the expense of safety by designing self-driving cars to zip through small gaps in oncoming traffic. Other situations that exemplify this mobility–safety tradeoff include merging onto highway lanes, driving through crosswalks with limited visibility, or avoiding detours by routing through school zones. Such maneuvers are very common, and we will soon be able to regulate them centrally via software when cars become more automated.

How should we make this trade-off between mobility and safety? What is the right balance? These are hard questions about the values that we build into self-driving cars. The ongoing debate on the ethics of self-driving cars typically focuses on two approaches to answering such questions: moral philosophy and social science. Although each has its place, both approaches fall short.

The first approach turns to moral theories. For example, a utilitarian theory that seeks to maximize overall happiness would probably recommend left turns, assuming that getting to your destination quickly creates more happiness than driving around the block to avoid left turns. But answering value questions purely by deduction from moral theories is a bad idea. First of all, it is an open question whether one moral theory is correct.[1] And even if we had one correct or uniquely best supported moral theory, we still should not rush to implement the answer that this theory gives. We value human agency and individual autonomy. People should be free to do what they believe is right even if that involves making certain mistakes [1].

The second way of answering ethical questions turns to psychology and the social sciences. We could gather data on how people in fact make this trade-off between mobility and safety. We could observe how people drive today; or we could survey people on how they think self-driving cars should operate; or we could look to how much people value mobility and safety in other forms of transportation such as air travel, and use this value elicitation to inform the balance between safety and mobility for self-driving cars.

---

[1] This is not to say that any moral theory is as good as the next one. You can believe that some theories rest on better arguments than others.

But, although understanding all relevant empirical data is crucial, relying exclusively on empirical facts is also a bad idea. People often reveal discriminatory or unfair preferences. For example, how likely a driver is to yield at a crosswalk may differ with a pedestrians' age, race and gender [2]–[4]. Although such data are illuminating and informative, we should be hesitant to build such preferences into our design.

Even if people had fair and unbiased preferences, insights from social science would still not be sufficient. First, although social science is useful for surfacing and quantifying disagreements, it provides only limited guidance on resolving disagreements. How we should solve disagreements is itself a moral question about which people will disagree. Second, people are to some extent self-interested, which creates a compliance or incentive problem. For example, if you are driving in a self-driving car you might want the car to protect your own safety at the expense of the safety of others [5]. Sometimes you should be allowed to prioritize your own interests to some extent and not put what you believe is morally right for others before what you believe is best for yourself. The balance between mobility and safety needs to be struck in a way that is compatible with individuals' fair pursuit of self-interest. This raises the questions: To what extent should we control, compel or force people to do what they themselves believe to be morally correct? Is the majority endorsement of a policy enough to force everyone into compliance? We should not answer these questions by majority opinion.

In sum, the two approaches of answering questions about the ethics of self-driving cars are both lacking. We should neither deduce answers from individual moral theories nor should we expect social science to give us complete answers. What further ways of addressing ethical questions can we draw on?

I argue that we should turn to political philosophy in addition to moral philosophy. The issues we face are collective decisions that we make together rather than individual decisions we make in light of what we each have reason to value [6]. A basic mistake in the ethics of self-driving cars is asking only what an individual should do. This is the domain of moral philosophy. Whether you should eat meat or maintain a vegetarian diet is an example of a moral question. But in addition to such questions, we also need to ask what makes for good policies and institutions. Policies and institutions result from collective decisions and form the domain of political philosophy.

Political philosophy adds three basic concerns to our conceptual toolkit.

First, political philosophy starts with the idea of reasonable pluralism, or the recognition that people often disagree for good reasons [7]. Almost all issues are complex, and a large amount of considerations and evidence can be brought to bear on the issues. At the same time, every one of us has different life experiences that inform our values. We might disagree for good reasons – perhaps we should cherish such a pluralism of values.

Second, political philosophy needs to balance values with a respect for human agency and individual autonomy. Even if everyone agreed that eating meat is morally bad, we would still need to ask how eating meat should be regulated. Within reasonable limits, individuals should be free to decide for themselves. We should not confuse questions on the individual level – What is wrong with eating meat? How should your self-driving car drive? – with questions on the collective level: How should we regulate meat consumption? How should we regulate self-driving cars?

Third, political philosophy worries about legitimate authority and why we are obliged to abide by decisions of our political institutions. What is regulated, how it is decided and by whom makes a difference. For example, even if we all agreed that eating meat is wrong, we might still hold that a private company lacks the authority to regulate meat consumption. Maybe certain issues need to be left unregulated. Maybe certain things should be done only by the government. These are all issues of legitimate authority.

These three concerns – reasonable pluralism, individual agency, and legitimate authority – are central themes in political philosophy. They have so far been largely overlooked in the debate on the ethics of self-driving cars. To illustrate this idea, think of the cases that have captivated the public discussion on the ethics of self-driving cars [8]. These so-called trolley cases ask you to imagine that a self-driving car needs to choose who has to die. If a car has a choice between running over five pedestrians or only one, when all pedestrians are identical in all relevant respects, what should the car do?

A major problem with such trolley cases and other such dilemmas is that they look at these choices as if they were exclusively a moral problem even though they raise a distinctively political problem. Trolley cases ask: What is the right thing to do? What would you do? What should the car do? But instead we need to think more broadly about value pluralism, individual agency, and political legitimacy when developing self-driving cars.

Self-driving cars — whether it is about trolley cases or left turns — raise the question of how we get along as a community or people. We now have a chance to regulate traffic systems to a degree that was just technically impossible before. When we think about good regulation, we can import concepts from political philosophy to inform our collective decision-making. For example, letting passengers of self-driving cars set at least some of the driving parameters themselves would be one way of achieving greater respect for reasonable pluralism, individual autonomy, and legitimacy.

This point generalizes to other issues, such as the ethics of artificial intelligence. How do recently heralded AI ethics principles respect reasonable pluralism, individual agency, and legitimate authority? How do corporate ethics principles reflect the diversity of values in society?

To respect reasonable pluralism, our responses to ethical challenges of technology should find common ground between the fundamentally different values that people hold and consider processes of incorporating and resolving value conflicts [9].

To respect individual agency, we should hesitate to replace human judgement and decision-making. And if we make decisions that affect others, we should keep their interests in mind and try to decide for them as they would for themselves.

To respect legitimacy, corporations should democratize decisions that affect basic goods such as individuals' liberty, safety, or opportunity. Options to democratize corporate decisions can involve government regulation, stakeholder participation, or deliberative input from the public, among others.

How such political concerns bottom out in policies in detail is a question rife for interdisciplinary cooperation. We should no longer overlook political philosophy when deliberating about the social challenges that arise from new technologies but make pluralism, agency, and legitimacy central pillars of the discussions.